# DiMA: Sequence Diversity Dynamics Analyser for Viruses

Shan Tharanga[1], Eyyüb Selim Ünlü[2], Yongli Hu[1,3], Muhammad Farhan Sjaugi[1], Muhammet A. Çelik[4], Hilal Hekimoğlu[4], Olivo Miotto[5,6], Muhammed Miran Öncel[4], and Asif M. Khan[1,4,*]

[1] Centre for Bioinformatics, School of Data Sciences, Perdana University, Kuala Lumpur, 50490, Malaysia
[2] Istanbul Faculty of Medicine, Istanbul University, Istanbul, 34093, Türkiye
[3] Department of Biochemistry, Yong Loo Lin School of Medicine, National University of Singapore, 8 Medical Drive, Singapore 117597, Singapore
[4] Beykoz Institute of Life Sciences and Biotechnology, Bezmialem Vakif University, 34820, Türkiye
[5] Nuffield Department of Clinical Medicine, University of Oxford, Oxford, OX3 7LF, U.K.
[6] Mahidol-Oxford Tropical Medicine Research Unit, Faculty of Tropical Medicine, Mahidol University, Bangkok, 10400, Thailand.

* To whom correspondence should be addressed: asif.khan@udst.edu.qa

Present Address:

Eyyüb Selim Ünlü: Wellcome Sanger Institute, Hinxton, UK
Yongli Hu: Beyond Limits SG Pte Ltd, 13 Stamford Road, Capitol Singapore, 178905, Singapore
Muhammet Celik: Biomedical Center (BMC), Physiological Chemistry, Faculty of Medicine, LMU Munich, Planegg-Martinsried, Germany
Asif M. Khan: College of Computing and Information Technology, University of Doha for Science and Technology, Doha, Qatar

## Abstract

Sequence diversity is one of the major challenges in the design of diagnostic, prophylactic and therapeutic interventions against viruses. DiMA is a novel tool that is big data-ready and designed to facilitate the dissection of sequence diversity dynamics for viruses. DiMA stands out from other diversity analysis tools by offering various unique features. DiMA provides a quantitative overview of sequence (nucleotide/protein) diversity by use of Shannon's entropy corrected for size bias, applied via a user-defined *k-mer* sliding window to an input alignment file, and each *k-mer* position is dissected to various diversity motifs. The motifs are defined based on the probability of distinct sequences at a given *k-mer* position, whereby an index is the predominant sequence, while all the others are (total) variants to the index. The total variants are sub-classified into the major (most common) variant, minor variants (occurring more than once and of frequency lower than the major), and the unique (singleton) variants. DiMA allows user-defined, sequence metadata enrichment for analyses of the motifs. The application of DiMA was demonstrated for the alignment data of the relatively conserved Spike protein (2,106,985 sequences) of severe acute respiratory syndrome coronavirus 2 (SARS-CoV-2) and the relatively highly diverse Pol protein (3,874) of human immunodeficiency virus-1 (HIV-1). The tool is publicly available as a web server (https://dima.bezmialem.edu.tr), as a Python library (via PyPi) and as a command line client (Via GitHub).

**Keywords:** viruses; proteome; genome; diversity; dynamics; sequence, motifs, conservation, entropy.

## Introduction

Viral infectious diseases are a major public health threat. There are more than 200 viruses that are considered infectious to humans (*1*). Viral infections have the potential to be of pandemic proportion, as demonstrated by the severe acute respiratory syndrome coronavirus 2 (SARS-CoV-2), responsible for the recent global health crisis, COVID-19. The first infection was reported in Wuhan, China, which eventually spread to the world at large, infecting more than 774 million people, with over seven million deaths (as of February 2024; https://data.who.int/dashboards/covid19). There is a need to better understand viruses to help mitigate these devastating effects, in particular for pathogens that are of global priority, such as *Filoviruses and Flaviviruses,* among others (https://www.niaid.nih.gov/research/emerging-infectious-diseases-pathogen).

The SARS-CoV-2 pandemic exemplified the value of sequence data in developing intervention strategies against viruses (*2*), such as diagnostics, therapeutics, and prophylactics. Sequence data is a treasure trove to better understand viral evolution and interaction with the host. It allows the study of viral diversity, which can help identify the changes and evolutionary pressures. Sequence change of even a single amino acid can affect the recognition of a virus by the host immune system (*3*). Viral diversity is a result of sequence substitutions that are primarily an outcome of evolutionary forces, such as mutation, recombination, and/or reassortment (*4*). Among these, mutations have the most significant effect, the rate of which can range from $10^{-8}$ to $10^{-6}$ and $10^{-6}$ to $10^{-4}$ nucleotide sites, per cell infection for DNA and RNA viruses, respectively (*5*). The variation within a single host can result in a spectrum of viral variants that are described as viral quasispecies (*6*).

Various studies have looked into the analyses of viral sequence variability and conservation. This has been facilitated primarily by alignment-dependent and to some extent alignment-free approaches. The latter is used when data is big, or the diversity is prohibitively large to allow for a reliable alignment (*7*). This approach, however, is not applicable if the goal is to study site-dependent substitutions (*8*). An alignment-dependent approach allows us to study conservation and variability across the sequences of a viral species and enables comparative analysis across related species. There are various alignment-based approaches to study diversity, such as through sequence similarity search (*9*), phylogeny (*10*), and biological patterns and profiles (*11*), among others.

A continuing goal is a greater understanding of viral diversity and an effective strategy to overcome the diversity for intervention applications (*12*). There is a need for a more detailed and quantitative analysis of the extent of viral substitutions, including the composition and incidence of the different variants of the viral genome/proteome (*13, 14*). Towards this, Hu *et al*. (*15*) presented a novel approach to dissect the dynamics of viral variability by use of diversity motifs for human immunodeficiency virus (HIV)-1 clade B. This was subsequently expanded and applied by Abd Raman et al. (*16*) to the influenza A (H5N1) virus. The authors used Shannon's entropy (*17, 18*), determined for overlapping aligned peptides of length *k*, via a sliding window, for a panoramic overview of the viral proteome diversity landscape, and further, as a novelty, each *k-mer* position was dissected to various diversity motifs. The motifs quantify the diversity dynamics for each of the overlapping *k-mer* positions by evaluating the incidence of the distinct sequences present at the position, which were classified broadly into four diversity motifs: index, major variant, minor variants, and unique variants (**Figure 1A**). The index is the predominant sequence at a given aligned position. It represents the prevalent wild-type sequence and if highly conserved and virus-specific, may be an attractive target for a vaccine, drug, and diagnostic designs. The remaining ranked sequences at the position are variants to the index, with at least one amino acid (or nucleotide) difference. The major variant is the second most predominant sequence. The minor variants are distinct sequences that occur more than once and are of frequency (incidence) lower than the major variant. Unique variants are observed only once in the alignment.

Herein, we present DiMA (Diversity Motif Analyser), a tool designed to facilitate the quantification and dissection of sequence diversity dynamics for any virus. It is publicly available both as a web server (https://dima.bezmialem.edu.tr) and as a standalone command-line interface (CLI), client tool (Python Package Index: https://pypi.org/project/dima-cli), useful for datasets of size beyond the web server limit. DiMA is big data ready, allows for user-defined sequence metadata enrichment, and enables comparative protein and nucleotide (RNA/DNA) sequence diversity dynamics analyses, within and between sequences of viral species. While the entropy function is also offered by various other tools, the novel key distinguishing feature of DiMA is the use of diversity motifs to study sequence diversity dynamics, besides integrating entropy and metadata.

## Materials and Methods

### Overall Workflow

DiMA workflow is illustrated in **Figure 1B**. The workflow can be divided into three parts, input, process, and output. The input to DiMA is simply a multiple sequence alignment (MSA) file of either protein or nucleotide sequences (in aligned FASTA format). The sequences can be obtained from publicly available primary/derived databases, such as NCBI Nucleotide/Protein/Virus databases (*19*) and ViPR (*20*), or specialist databases, such as Influenza Research Database (IRD) (*21*), among others. Both full-length and partial sequences are recommended for a comprehensive survey of diversity. Primary databases are prone to errors (*22-24*), and thus derived databases, which are integrated, value-added resources, help minimize these issues. NCBI Virus is recommended for its ease of use and data quality; nonetheless, sequence data records of interest should be reviewed for any irrelevant sequences and possible issues, such as errors, discrepancies, and anomalies (*25*). Sequence duplicates (full-length or partial match) are common in public databases, which may be an artifact (multiple uploads of the same isolate) or an actual indication of virus incidence in nature (multiple uploads of different isolates with the same sequence). Thus, it may be desired to analyse both the redundant and non-redundant datasets (*26*).

There are many tools available for MSA, such as Muscle (*27*), Clustal Omega (*28*), and MAFFT (*29*), among others. MAGUS (*30, 31*) is suitable for big data (up to one million sequences). Given the less optimal, heuristic nature of MSA (*32-34*), all resulting alignments should be manually inspected and corrected for any misalignments; short partial sequences can be a common cause of spurious alignments. The tool EMBOSS Seqret (https://www.ebi.ac.uk/jdispatcher/sfc/emboss_seqret) may be used to convert an aligned file to an aligned FASTA format for use in DiMA. DiMA provides a quantitative measure of sequence diversity by use of Shannon's entropy, applied via a user-defined *k-mer* sliding window on the alignment. The entropy value is corrected for sample size bias by applying a statistical adjustment. DiMA interrogates the diversity dynamics by dissecting each *k-mer* position to various diversity motifs. Collectively, DiMA outputs plots of the entropy values, diversity motifs, and user-enriched metadata (such as year, host, and country) for each of the *k-mer* positions, providing a holistic view of the diversity and its dynamics.

**Input and various considerations**

The input MSA file (up to 100 MB; larger files are possible with the CLI version) is accepted only in an aligned FASTA format (sample input files are provided on the web server), and the analysis can be customized through various parameter settings and features:

*k-mer length*

Genetic diversity is defined as changes at the nucleotide level. However, not all genetic changes result in changes at the protein level; only non-synonymous substitutions give rise to different amino acids. Furthermore, not all amino acid changes are antigenic and affect recognition by the host immune system. For example, a single amino acid change may not abrogate the immune recognition of a known epitope or give rise to a new epitope. Antigenic changes are more punctuated than protein sequence changes (*35*). Thus, it is important to study the genetic, protein, and antigenic sequence diversities of viruses.

A sliding window *k-mer* approach is better suited to capture the local (neighbouring) effect of substitutions, in particular when dealing with longer sequences (*36*). Although a window size of one (1) is acceptable for a diversity study, the various considerations for larger *k-mer* sizes are discussed in Chong *et al*. (*7*), albeit, largely in the context of alignment-free comparisons. With regards to antigenic diversity, a *k-mer* length of nine (9; nonamer) has been used to study antigenic diversity in the context of cellular immune response. This is because the human leukocyte antigen (HLA)-I molecules can bind peptides of lengths ranging from eight to 15 amino acids (aa), with nine aa being the typical length (*37*) and HLA class II molecules can bind longer peptides, up to 22 aa, with a binding core of nine aa (*38*), (*26, 39*).

DiMA employs a user-defined *k-mer* sliding window to measure the entropy of the aligned sequences in the input file and perform the subsequent diversity dynamics analyses. While the minimum size is one, the maximum can equal the alignment length of the input file. By default, DiMA uses a window size of nine amino acids and 27 nucleotides for protein and DNA alignments, respectively. However, the user may specify any desired window size within the alignment length limit.

*Metadata Parsing*

The description/definition line or headers of the input-aligned sequences in FASTA format can be utilized to annotate the sequence with meaningful metadata, which can provide additional dimensions to viral diversity studies. Such metadata are often available in public databases or could be from in-house findings. The NCBI Virus database, for example, provides up to as many as 23 standard metadata annotations to each sequence record, including host, biosample, geolocation, and year of isolation, among others. The input file can be formatted to include metadata as desired by the user, whether taking advantage of information from the public database record and/or from in-house. DiMA provides a feature to parse and tag such metadata to the sequence headers of the input alignment file. It currently provides six pre-defined, commonly used header tags, such as “Accession”, “Strain”, “Species”, “Year”, “Geolocation”, and “Host”; and users can also define their own custom tags. The pipe ("|”) character is used as a delimiter between the tags, while the order of the metadata in the input file and those of the tags should match. As the number of metadata available can vary between sequences in an input file and since the order is important, empty pipes ("||”) to reflect unknown/unavailable metadata, must be added to preserve the order of the header tags. Two examples of sequence headers are “AGN52936.1 |Homo sapiens|United Arab Emirates|2013” and “>YP_009047204.1 |Homo sapiens||2012”, which correspond to header tags as follows: “Accession”, “Host”, “Geolocation”, and “Year”. The NCBI Virus database allows customization of the FASTA header line, with the inclusion of empty pipes where applicable. The user-defined header tags in DiMA would be accordingly ascribed to all the *k-mers* generated from the input-aligned sequences.

### *Support Threshold*

The number of sequences in an input alignment file affects the reliability of insights from the data. Ideally, large sample size is preferred to mitigate the effect of any potential data biases. Herein, “support” is defined as the number of sequences at a given *k-mer* position that do not harbour a gap and/or an unknown/ambiguous nucleotide base or amino acid residue. The *k-mer* positions that are below a user-defined “support” threshold (default is 30, arbitrarily defined) are referred to as “low support” (tagged as “LS”). A position of “no support” (tagged as “NS”) is possible when all of the aligned sequences at a given *k-mer* position have a combination of gap(s) and/or unknown/ambiguous character(s). Support may vary between *k-mer* positions due to the inclusion of incomplete or partial sequences into the alignment.

## Process

### *Shannon’s entropy*

Shannon’s entropy, originally introduced as a theory of communication, is used to measure the disorder for a given variable and has been widely adopted in biology as a diversity index, including for sequences (*26, 39*). DiMA calculates the entropy, *H(x)* for each overlapping, *k-mer* position *x* of the input alignment file by applying Shannon’s formula:

$$H(x) = -\sum_{i=1}^{n(x)} p(i,x) \log_2 p(i,x)$$

where *p(i,x)* is the relative frequency (probability) of a given *k-mer* sequence *i* at the position *x* (start position of the *k-mer*) and *n(x)* is the number of distinct *k-mer* sequences at the position. These two factors affect the entropy value, whereby it is high for a position with a large number of distinct *k-mer* sequences and is low when a sequence exhibits a clear high probability. The minimum possible entropy value is 0, which means complete conservation of a given *k-mer* position, whereby one *k-mer* sequence is observed in all (100%) of the aligned sequences analysed. The maximum entropy value would be dependent on the *k-mer* size of choice and is applicable when all possible outcomes (permutations with repetition) are observed with equal probability. The maximum peptide entropy value for a *k-mer* window size of nine is 39 (i.e., $\log_2 20^9$). This, however, is theoretical for biological sequences given that conservation is expected among homologs, and some combination of *k-mers* may not be sterically possible. The highest nonamer entropy value that we have thus far reported is 9.2 for HIV-1 clade B envelope protein (*15*), and it may be used as a benchmark for comparative proteome diversity dynamics analyses.

The entropy computation for a given *k-mer* position is dependent on the number of sequences in the input alignment. Only *k-mer* sequences that do not harbour a gap and/or unknown/ambiguous nucleotide base or amino acid residue are used for the entropy computation. Entropy is calculated irrespective of the support tag ("LS" or "NS") for a position. NS positions are maintained to keep the input alignment length intact, with a default assigned pseudo entropy value of zero, which is cautioned with the inclusion of the NS tag in the results page and the downloadable output files.

Raw entropy values for each *k-mer* position are corrected for sample size bias (*40*) by applying a statistical adjustment to the input alignment, which estimates entropy values for infinitely sized resampled alignments with analogous *k-mer* distribution (*26*). Briefly, this is done by resampling the sequences that can be analysed ($N$) at a *k-mer* position and creating alignments of varying size, determined through a systematic random sampling approach. The entropy values are measured for each size, plotted against the respective $1/N$, and a linear regression is used to extrapolate the entropy estimate for the *k-mer* position when $N \to \infty$. Resampling for an infinite-size set estimate is only done for positions when N is higher than the support threshold ($T$); otherwise, all of $N$ is used for a direct computation of the entropy. An exceptional low support ("ELS") tag is used in the results page and downloadable output files when $N = T$, as a way to distinguish from when $N < T$. In the case of N only slightly higher than T and where regression may potentially result in a negative entropy value due to small number of resamples, all of N is used for a direct compute of the entropy. The entropy calculation algorithm is illustrated on the DiMA manual page (https://dima.readthedocs.io/en/latest/#entropy-algorithm).

*Motif classification*

DiMA interrogates the diversity dynamics at each *k-mer* position by classifying the distinct sequences into diversity motifs, based on their frequency (probability) (**Figure 1A**). The most frequent *k-mer* sequence is termed as the "index", with "major variant" as the second most common and "unique variants" as the least, occurring only once, while "minor variants" are in between these two variant motifs. As per the motif definition, certain *k-mer* positions may not exhibit specific motifs, such as no index, major or minor, and only singletons are observed. Also, in some instances, a position may exhibit more than one index or major variant, when two or more distinct *k-mer* sequences are of the same frequency. The term "total variants" encompasses all the variants that are at least one nucleotide/amino acid different from the index. The term "distinct variants" is the count of the different *k-mer* sequences among the total variants (major, minor, and unique variants). The relative frequency of each motif's distinct sequences, expressed in percentage (as incidence) is computed by DiMA for each of the *k-mer* positions. The denominator used for the incidence computes is $N$ (for each of the motif's distinct sequences), except for the term "distinct variants", where the total variants count is used. As such, a value of 100% for "distinct variants" will only be attained when all the *k-mers* are unique variants (no major and/or minor variants). Additionally, DiMA identifies historically conserved sequences (HCSs) across the input alignment length, selected based on a user-defined index incidence threshold (default: 100%). Selected index sequences that overlap or are immediately adjacent in their *k-mer* positions are concatenated as longer sequences. An HCS of 80% or higher incidence may be attractive for further consideration as an intervention target.

### *Stratification by metadata*

At each *k-mer* position, an individual distinct sequence is populated with the corresponding metadata header tags from all the input alignment sequences that share the same distinct sequence. Accordingly, the relative frequency for each of the different constituent items of each header tag is calculated and plotted.

## **Output**

### *Interactive interface*

DiMA outputs a result page with 12 different panels, organised into three parts (top, middle and bottom) that offer multiple facets of sequence diversity (**Figure 1C**). The multiple panels at the top are akin to a dashboard that provides the user with a quick summary of information general to the input alignment and specific to a given *k-mer* position: i) alignment length, ii) download results, iii) query name, iv) support threshold, v) position support, vi) distinct variants, vii) position entropy, and viii) selected position. The single middle panel (ix) showcases the entropy plot for all the *k-mer* positions of the input alignment, providing an overview of the sequence diversity, with low entropy values representing conservation and high indicating variability; hovering the cursor over a specific position on the plot will reveal the applicable entropy value. The three panels at the bottom provide information for the selected position, namely x) motif distribution and xi) *k-mer* position distinct sequences, whereby the selection of a specific distinct *k-mer* sequence would illustrate all the relevant xii) sequence metadata, with a drop-down menu option for the different header tags defined by the user. Changing the selected *k-mer* position (in panel viii) would dynamically update the information in panels (x to xii).

### *Download*

The top panel (ii) of the DiMA output allows for downloading of the analyses results in JSON and XLSX formats. The JSON file provides the complete analyses results as key-value pairs, which can be viewed using a public JSON viewer tool (such as https://jsonformatter.org/json-viewer). Separately, the XLSX file provides for easier viewing through the MS Office Excel application or an equivalent that supports the format. A concatenated list of HCS based on a user-defined index incidence threshold can also be downloaded (in JSON format) and viewed in a text editor or JSON viewer.

## **DiMA implementation**

DiMA is designed using the Python language, with bindings written in Rust. This allows DiMA to remain highly performant while being accessible to the larger Python user base. Designed with big data in mind, DiMA can handle large datasets while still maintaining a relatively small memory footprint. Furthermore, DiMA implements data parallelism where each *k-mer* position is handled in parallel, making it efficient and scalable. In addition, the dataset uploaded by the user on the web service is validated for MSA and header consistency using functions written in Rust, compiled down to WebAssembly. This allows for efficient validation of large datasets and mitigates user's submission from failing once sent to our server for processing. DiMA is compatible with common web browsers, including Chrome 98.0, Firefox 56.0, Edge 94.0, and Safari 15.0.

## Results

## Application

We demonstrate two applications on the use of the DiMA web server for viral sequence diversity dynamics studies. One is using the Spike protein of *severe acute respiratory syndrome coronavirus-2* (SARS-CoV-2), with raw sequence data and metadata downloaded from the GISAID (*41*) database. The second is the DNA polymerase (Pol) protein of human immunodeficiency virus type 1 (HIV-1) group M, with curated alignment sourced from the HIV Sequence Compendium 2021 (42), made available on the Los Alamos HIV sequence database (*41*). SARS-CoV-2 Spike protein plays a pivotal role in host cell entry by binding to angiotensin-converting enzyme 2 (ACE2) receptors, and thus is a key target for developing therapeutics to block viral entry (*42*). HIV-1 Pol protein provides essential enzymatic functions critical for viral replication and has been a major target for drug design (*43, 44*).

All 11,808,363 available Spike protein sequence records were downloaded (as of July 2022) and deduplicated by use of CD-HIT v4.8.1 (*45*). The resulting 2,106,985 non-redundant sequences were multiple sequence aligned by use of an in-house reference guided MSA tool, RefAligner (using the GISAID accession ID "EPI_ISL_402124" sequence as a reference). The aligned dataset was analysed using the command-line version of DiMA (DiMA-CLI), which enables big data analysis, with metadata parsing enabled and minimum incidence for HCS concatenation set to 99.95, while other parameters set to the default. The analysis results are made available on the DiMA web server (https://dima.bezmialem.edu.tr/results/6cfb645e-eefa-40db-ac4c-43a6712b2760). A total of 1,265 aligned nonamer (9-mer) positions were analysed and the protein was generally conserved (**Figure 2A**), with an average entropy of ~0.28. Position 675 *(the number refers to the starting alignment position of the k-mer)* peaked as the most diverse position, with the highest entropy value of ~1.97. This position had a total support of 2,208,211 sequences, with 343 distinct 9-mers. The predominant distinct 9-mer (`QTQTNSRRR`) or the index at this position occurred in 876,048 of the sequences analysed, and thus, exhibited an incidence of ~39.67%. The remaining 342 nonamers at the position were variants to the index and occurred across the 1,332,163 remaining sequences analysed. The total variants (~60.33%) of the position comprised of a major variant (`QTQTKSHRR`; 833,415; ~37.74%), 234 distinct minor variants (~22.58%), and 108 unique variants (~0.005%). Among the total variants, the total number of "distinct variants" was 343 (~0.03%). Sequence metadata visualization showed that the index was observed primarily in *Homo sapiens* (~99.84%), with the remaining (~0.16%) in other host species such as *Canis lupus familiaris*, *Panthare leo*, *Neovison vison*, *Fells catus, Odocoileus virginianus*, *Mustela lutreola* and environmental sources. While the index was found in 29 countries, it was most commonly detected in two geographical locations: United States (~35.33%) and United Kingdom (~13.73%). The index 9-mer was most common among sequences collected in January 2022 (~13.10%), and with decreased frequency over the following three months (~7.46, ~4.82, and ~3.91). DiMA identified five historically conserved sequences (HCSs) of lengths ranging from 9 to 23 amino acids, which merit further investigation as potential targets for vaccine design, inhibitory drug analysis, and diagnostic studies.

No data processing was done for the Pol protein as a pre-curated alignment (2,472 sequences; as of 2021) was retrieved; however, did not include any metadata. DiMA analysis of the protein was performed with the following parameters: support threshold of 100 to discard positions of alignment reliability, HCS threshold arbitrarily set to 99.95% (none was observed at 100%), and metadata parsing disabled. The analysis result is available here (https://dima.bezmialem.edu.tr/results/c5cbae80-6a36-43ef-928a-4d04d8044f06). A total of 1,189 aligned 9-mer positions were analysed, with 266 (~22.37%) identified as low support (LS) and 436 (~36.67%) of no support (NS). Ignoring LS and NS positions, the protein was relatively more diverse than SARS-CoV-2 Spike. The Pol entropy values ranged from 0.0 to ~6.19, with an average of ~0.84 **(Figure 2B)**. The position with the highest entropy only had a support of 34. Position 772 had the next highest entropy of ~5.75, but not the highest total variants (~81.17%), which was exhibited by position 225 (total variants of ~85.29%) (**Figure 2C**), illustrating the complex composition of the viral variant population. The index 9-mer, `IPIEICGHK`, at position 225 was only observed at an incidence of ~14.71%, while it was reported at ~30% in Hu *et al*. (*15*) of data from 2008, indicating a further diversification of total variants, with the major variant remaining relatively stable (only a ~1% incidence increase). The minor variants were instead the principal variant with a collective incidence of ~66.5% at the position. The lowest absolute total variants incidence of ~0.69% was observed at the highly conserved position 433 (**Figure 2D**), with an entropy of ~0.08. In contrast to position 225, the incidence of the index (`NVLPQGWKG`) at 433 was ~99.3%, with negligible total variants (~0.7%). Among the variants, no distinct 9-mer was dominant and thus, a high "distinct variants" incidence (~61.5%) was observed.

The above are cursory evaluations of the two proteins. Data from DiMA can be synthesised and analysed further in various ways, such as i) scatter plot of the relationship between entropy and incidence (% frequency) of total variants; ii) scatter plot of motif incidence (for each diversity motif) against total variants; iii) incidence distribution violin plots of the diversity motifs; and iv) distribution of conservation level of index incidence for *k-mer* positions. This can be done for individual proteins/genes of a virus and further pooling of the results can allow for a genome/proteome-wide analyses. Examples of such analyses are demonstrated In Hu *et al*. (2013) and Abd Raman *et al*. (2020). Further, a notable finding from these studies was motif switching, a phenomenon where the fitness change of one or more nucleic or amino acids, changed the incidence, and thus the motif of a given *k-mer* distinct sequence across its *k-mer* positions. Motif switching has been observed to involve all the diversity motifs studied herein. Hence, DiMA results also serve as a starting point to unravel and understand the complexity of motif switching.

**Discussion**

DiMA stands out from other diversity analysis tools, such as Protein Variability Server (PVS) (*46*) and Los Alamos (LANL) (http://www.hiv.lanl.gov/) virus databases Entropy tool, by offering various unique features. A comparison table between these tools and DiMA can be found in our help page (https://dima.readthedocs.io/en/latest/#novel-features) and also provided herein (**Table 1**). Notably, DiMA is capable of handling both nucleic acid and amino acid sequences. Shannon's entropy is calculated for a user-defined *k-mer* size of a sliding window, which is better suited to capture the local (neighbouring) effect of sequence substitutions, in particular when dealing with longer sequences. A statistical adjustment is applied to the computed entropy values for sample size bias correction. Distinctively, the key feature of DiMA is that it interrogates the diversity dynamics by dissecting each *k-mer* position to various diversity motifs, based on the probability of distinct sequences. Moreover, it allows for metadata enrichment of the motifs. Additionally, DiMA identifies historically conserved sequences (HCSs) across the input alignment length, selected based on a user-defined index incidence threshold, which are concatenated when overlapping or adjacent.

DiMA outputs a result page with 12 different panels, organised into three parts (top, middle and bottom) that offer multiple facets of sequence diversity and are largely interactive. The input to DiMA is simply a multiple sequence alignment file of size up to 100 MB, while larger files can be accommodated by use of the command line (CLI) version.

DiMA is a tool that enables sequence diversity dynamics analysis for any virus of interest. It is big data ready and applicable to both DNA/protein sequence data, with a user-friendly interface and a detailed user manual provided for a broader appeal. DiMA enables comparative sequence diversity dynamics analyses for a better understanding and insight within and between protein/DNA sequences of a virus species or genomes/proteomes of different viral species, whether at the genus, family, or higher lineage taxonomy rank, only limited by the feasibility of a reliable alignment. The diversity motifs represent inherent patterns in the organisation of the large number of sequences that facilitate cooperative virus fitness selection, whereby the spectrum of variants allow the virus population to explore changes in selection pressure for maximum reproductive fitness (*47*) and support long-term evolvability (*47, 48*). In some cases, there may be functional constraints (*49*), such as multiple host dependency limiting the virus population from a fitness peak (*16, 26*) or in contrast, the presence of extremely variable hotspots contributing to the general collapse of the early immunity or facilitating immune escape (*15*). Examining viral variants, leveraging on various metadata dimensions of interest, such as spatio-temporal, host, and clinical phenotypes, among others, may provide important key insights into the evolution of the virus.

Sequence diversity at highly variable positions appears to be embodied in minor and unique variants, which can exist as many different (distinct) sequences. DiMA can help elucidate variant sequence structure and incidence with increased total variants. Thus, provides a compendium of the possible spectrum of sequence variants for a virus of interest. The viruses can range from being highly conserved, such as West Nile virus (*50*) to extremely variable and highly plastic, such as HIV-1 (*15*). DiMA analyses of different viral species can provide a catalogue of HCS targets for a comparative and rational design of new intervention strategies. DiMA can potentially be used for non-viral pathogens, such as bacteria and fungi, among others.

*Limitations*

DiMA's effectiveness is dependent on the quality of the alignment, which is constrained by the heuristic nature of MSA. This process relies on the subjective manual inspection and correction of misalignments by the user. Additionally, DiMA currently focuses only on sequence diversity in the context of substitutions and does not consider other structural sequence variations, such as copy number changes. Shannon's entropy is just one of the many diversity metrics available; other metrics may be explored in future iterations of DiMA. The input file size (up to 100MB) limitation of the web server is overcome through the offering of the GUI limited, CLI version. Despite these limitations, DiMA is a significant step forward toward the study of viral sequence diversity dynamics.

**Acknowledgment:** We gratefully acknowledge all data contributors, *i.e.*, the authors and their originating laboratories responsible for obtaining the specimens, and their submitting laboratories for generating the genetic sequence and metadata and sharing via the GISAID Initiative and HIV LANL, on which this research is based. We thank all those who used and/or evaluated DiMA and/or its earlier iterations, directly or indirectly, during the research and development phase (https://dima.readthedocs.io/en/latest/#acknowledgement).

**Funding:** The computational resources and services used in this work were provided by Perdana University School of Data Sciences, Malaysia and Bezmialem Vakif University, Turkey. AMK was supported by Perdana University, Malaysia, Bezmialem Vakif University, Turkey, and The Scientific and Technological Research Council of Turkey (TÜBİTAK). This publication/paper has been produced benefiting from the 2232 International Fellowship for Outstanding Researchers Program of TÜBİTAK (Project No: 118C314). However, the entire responsibility of the publication/paper belongs to the owner of the publication/paper. The financial support received from TÜBİTAK does not mean that the content of the publication is approved in a scientific sense by TÜBİTAK.

**Author contributions**: Conceptualization: AMK, YH, OM. Methodology: AMK, ST, ESÜ, YH, OM, MMÖ. Implementation: ST, MFS, MAÇ, AMK, ESÜ. Application: ESÜ, MAÇ, AMK. Supervision: AMK, MFS. Writing—original draft: AMK, HH, ESÜ. Writing—review & editing: AMK, ESÜ, HH.

**Competing interests:** None declared.

**Data and materials availability:** DiMA web server is implemented in Rust and Python version 3.10. and is accessible through https://dima.bezmialem.edu.tr. It is also freely available as a standalone command-line interface (CLI) client tool at https://pypi.org/project/dima-cli and the source code is accessible on the GitHub repository at https://github.com/PU-SDS/DiMA. DiMA documentation is available at: https://dima.readthedocs.io/en/latest/.

The findings of this study are based on data associated with 2,583,449 sequences available on GISAID up to March 18th, 2022, and accessible at http://doi.org/10.55876/gis8.230207sb, and 24,985 sequences available on HIV LANL, accessible at https://www.hiv.lanl.gov/content/sequence/NEWALIGN/align.html.

## Tables

**Table 1.** Novel features of DiMA in comparison with other web servers for viral sequence variation analysis

| Features | DiMA[1] | PVS[2] | LANL[3] | BV-BRC[4] |
|---|---|---|---|---|
| Analysis of nucleic acid - amino acid sequences | ✓ - ✓ | ✕ - ✓ | ✓ - ✓ | ✓ - ✓ |
| Shannon Entropy on user-defined sliding window | ✓ | ✕ | ✓ | ✕ |
| Entropy correction for size bias | ✓ | ✕ | ✕ | ✕ |
| *k-mer*/variant frequency calculation | ✓ | ✕ | ✕ | ✓ |
| *k-mer* diversity motifs classification | ✓ | ✕ | ✕ | ✕ |
| Metadata inclusion | ✓ | ✕ | ✕ | ✕ |
| Identification of historically conserved sequences | ✓ | ✓ | ✕ | ✕ |
| Multiple interactive visualizations | ✓ | ✕ | ✕ | ✕ |
| Web service input size limit | 100 MB* | ~0.2 MB | Not defined | Not defined |

1. https://dima.bezmialem.edu.tr
2. http://imed.med.ucm.es/PVS/
3. https://www.hiv.lanl.gov/content/sequence/ENTROPY/entropy.html
4. https://www.bv-brc.org/app/MSA

* Analysis of larger files possible with CLI version (https://github.com/BVU-BILSAB/DiMA) which there is no limit.

# Figures

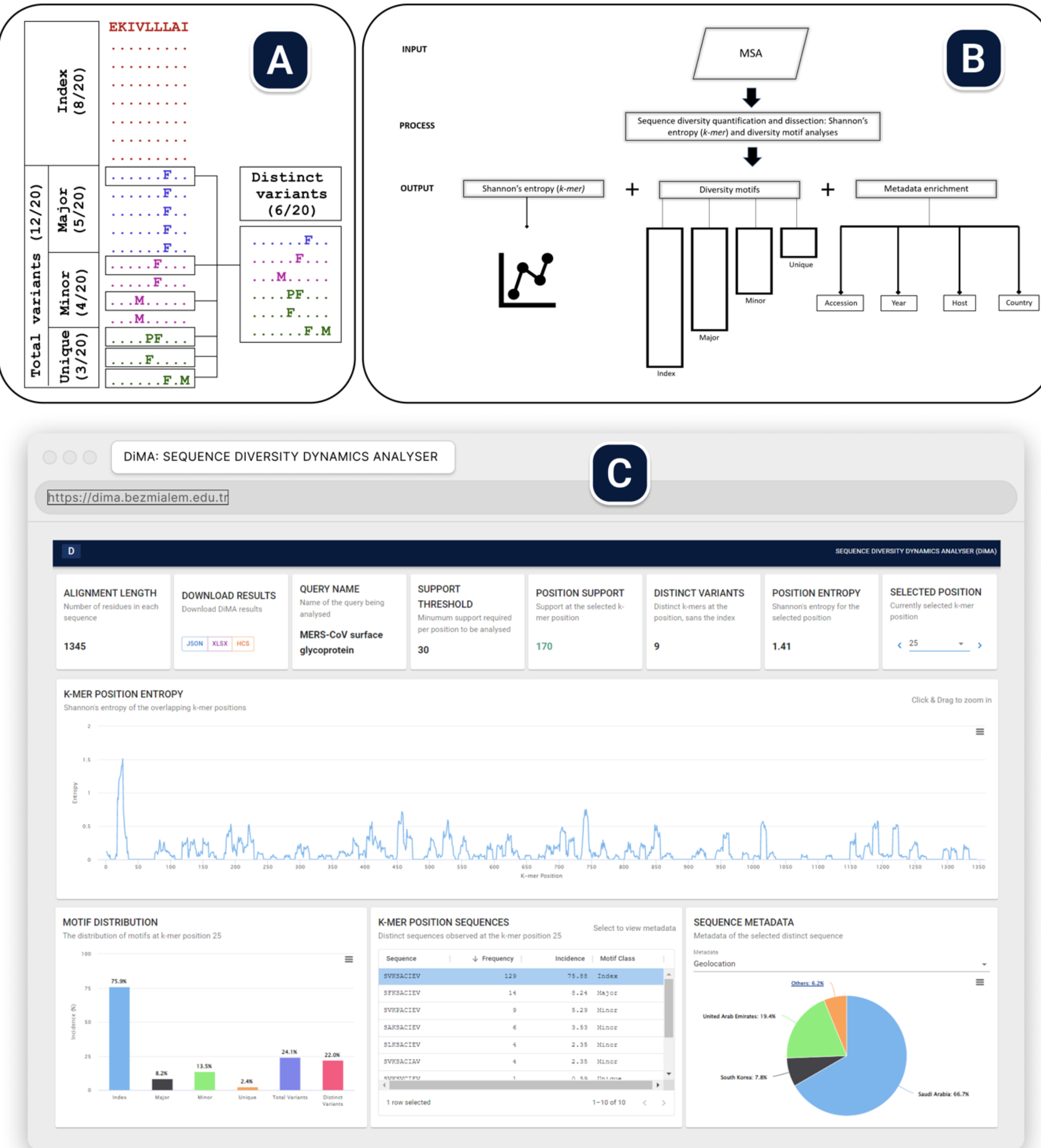

**Figure 1.** DiMA at a glance. **A.** Diversity motif definition for a sample *9-mer* position of 20 sequences. **B.** DiMA workflow. **C.** DiMA server results page. Submission summary (top panel): overview of information about the input alignment, such as the query data name, alignment length, the support threshold, and the support at the selected *k-mer* position, among others. *K-mer* position entropy (middle panel): entropy value of the selected *k-mer* position

indicates the level of variability among the sequences at the position, with zero representing a completely conserved position. The entropy plot provides a panoramic view of the diversity and is responsive and interactive. Motif distribution (bottom panel): all sequences at the selected aligned *k-mer* position are quantified for distinct sequences and are ranked classified into diversity motifs based on their incidences (probability). *K-mer* position sequences (bottom panel): distinct sequences at the selected aligned *k-mer* position are listed according to their corresponding incidences. Metadata (bottom panel): if the header tags are provided, DiMA will show a pie chart for each type of metadata.

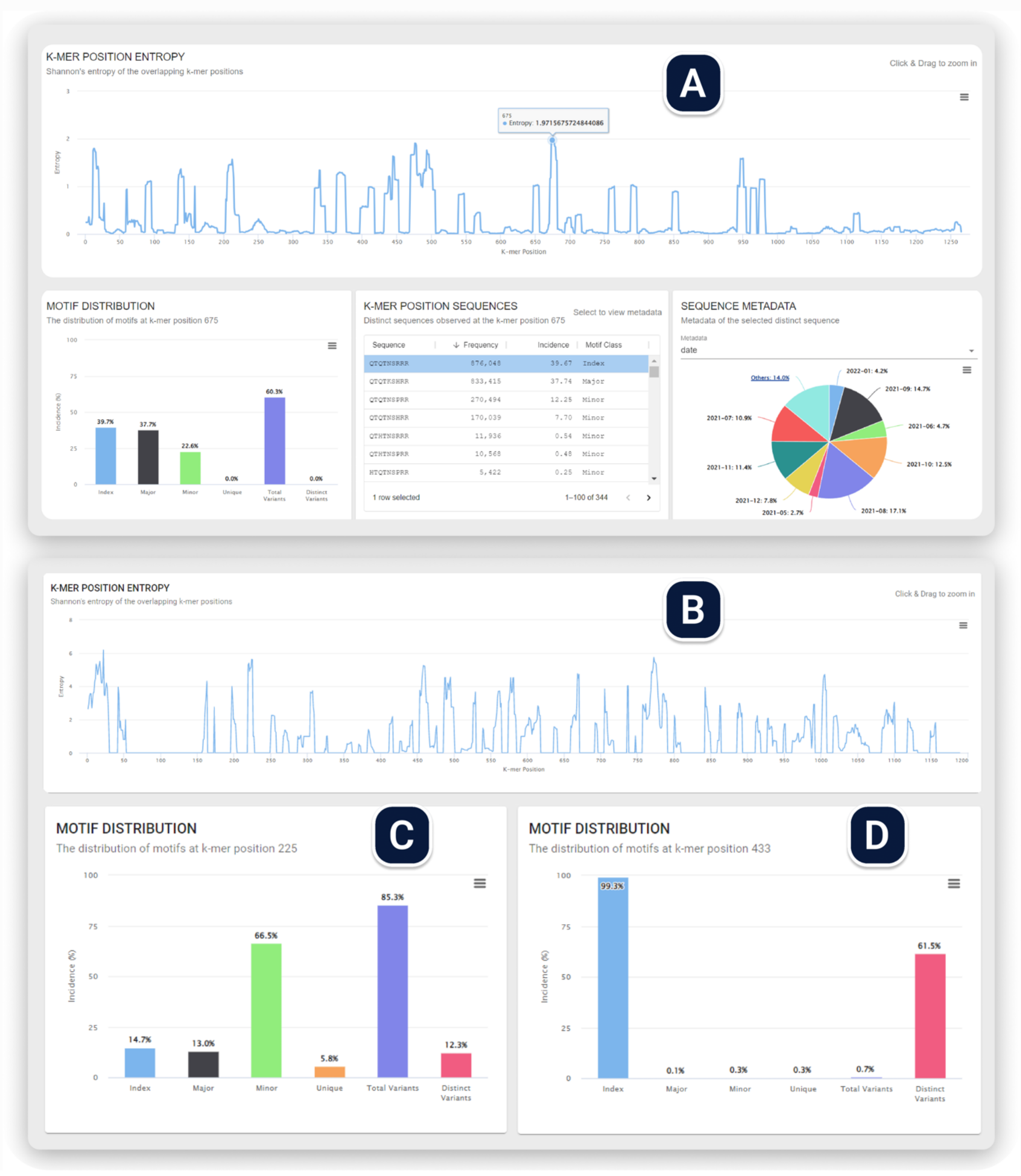

**Figure 2. Applications of DiMA for the diversity dynamics studies of SARS-CoV-2 Spike and HIV-1 Pol proteins. A**. The highest entropy observed was ~1.97 for SARS-CoV-2 Spike protein at position 675 *(the number refers to the starting alignment position of the k-*

*mer)*, which was selected for motif distribution and metadata view of each distinct *k-mer* at the position. **B**. HIV-1 Pol protein exhibited various conserved and highly diverse positions. **C-D**. Motif distributions of a highly diverse position (225; highest total variants) and a highly conserved position (433; lowest total variants), respectively, in HIV-1 Pol.